\documentclass[journal]{IEEEtran}

\ifCLASSINFOpdf
\else
   \usepackage[dvips]{graphicx}
\fi
\usepackage{url}
\usepackage{cite}
\usepackage{amsmath,amssymb,amsfonts}
\usepackage{algorithmic}
\usepackage{graphicx}
\usepackage{textcomp}
\usepackage{caption}
\usepackage{xcolor}
\usepackage{color}
\usepackage{hyperref}
\usepackage{multirow,multicol}
\usepackage{subcaption}
\usepackage{fancyhdr}

\usepackage{graphicx}

\begin{document}

\title{Speaker-Text Retrieval via Contrastive Learning}

\author{Xuechen Liu, Xin Wang, \IEEEmembership{Member, IEEE}, Erica Cooper, Xiaoxiao Miao, Junichi Yamagishi, \IEEEmembership{Member, IEEE}
% \thanks{This paragraph of the first footnote will contain the date on which you submitted your paper for review. It will also contain support information, including sponsor and financial support acknowledgment. For example, ``This work was supported in part by the U.S. Department of Commerce under Grant BS123456.'' }
\thanks{This project is supported by Google.}
\thanks{Xuechen Liu, Xin Wang, Erica Cooper, Xiaoxiao Miao and Junichi Yamagishi are with National Institute of Informatics, Tokyo 101-8430, Japan (e-mail: \{xuecliu, wangxin, ecooper, miaoxiaoxiao, jyamagis\}@nii.ac.jp).} 
% \thanks{Xiaoxiao Miao is now with Singapore Institute of Technology, 10 Dover Drive 133863, Singapore (e-mail: XIAOXIAO.MIAO@singaporetech.edu.sg).}.
}

\markboth{Journal of \LaTeX\ Class Files, Vol. 14, No. 8, August 2015}
{Shell \MakeLowercase{\textit{et al.}}: Bare Demo of IEEEtran.cls for IEEE Journals}
\maketitle

\begin{abstract}
In this study, we introduce a novel cross-modal retrieval task involving speaker descriptions and their corresponding audio samples. Utilizing pre-trained speaker and text encoders, we present a simple learning framework based on contrastive learning. Additionally, we explore the impact of incorporating speaker labels into the training process. Our findings establish the effectiveness of linking speaker and text information for the task for both English and Japanese languages, across diverse data configurations. Additional visual analysis unveils potential nuanced associations between speaker clustering and retrieval performance.
\end{abstract}

\begin{IEEEkeywords}
Speaker Retrieval, Contrastive Learning, Multi-Modal Learning.
% Enter key words or phrases in alphabetical order, separated by commas. For a list of suggested keywords, send a blank e-mail to keywords@ieee.org or visit \url{http://www.ieee.org/organizations/pubs/ani_prod/keywrd98.txt}
\end{IEEEkeywords}

\IEEEpeerreviewmaketitle

\section{Introduction}
\IEEEPARstart{T}{he} boost of big data and hardware capabilities unveils enormous quantities of data being generated and acquired for various applications, which needs to be stored and queried effectively. Consequently, the need for enhanced contextual search capabilities on those data has been surging. Active research has emerged on aligning and retrieving cross-modal information, such as between image and text \cite{image_text_alignment2015, clip}, video and text \cite{video_text_retrieval2015, video_text_retrieval2019}, and video and audio \cite{video_audio_retrieval2019}.

Cross-modal retrieval between audio and textual information has been an active area of research. The primary goal of such a task is to query audio clips of various lengths from a text description. Starting from short audio clips with various sound effects \cite{audio_text_retrieval2008}, more capable models for long-form audio clips have been proposed based on a Siamese \cite{siamese2016} architecture, where the retrieval is done via a shared embedding space from two aggregation modules. Those models are often trained with cosine similarity \cite{msft_clap2023} or contrastive loss \cite{laion_clap2023}. They come with databases such as AudioSet \cite{audioset} and CLOTHO \cite{clotho}, pre-trained audio encoders \cite{pann, htsat}, and language models \cite{bert}. However, the primary focus of the linking is between audio event samples and their descriptions, and the main downstream task is thus broad audio classification.

Meanwhile, deep learning has also improved speaker encoders via \emph{deep neural network} (DNN) and discriminative training objectives. Speaker embeddings extracted from DNN-based speaker encoders have proven to be strong and fruitful representations for tasks such as speaker identification and verification. A recent study \cite{probe_xvector2019} reveals that they can also be useful for detecting various attributes such as phoneme and speaking rate. 

In this paper, we present a preliminary study on linking the speaker-wise and textual information, forming a fine-grained, effective, and explainable search space. We migrate earlier ideas from cross-modal contrastive learning frameworks \cite{clip, laion_clap2023, msft_clap2023} and highlight our approach's effectiveness on linking cross-modal information via a simple and light-weight model.
%built solely on fully-connected (FC) layers. 
We primarily benchmark such functionality on the task of cross-modal retrieval, while the framework can be further utilized and generalized to perform speaker generation, speaker characterization, and speech synthesis.

\begin{figure}
    \centering
    \includegraphics[width=\linewidth]{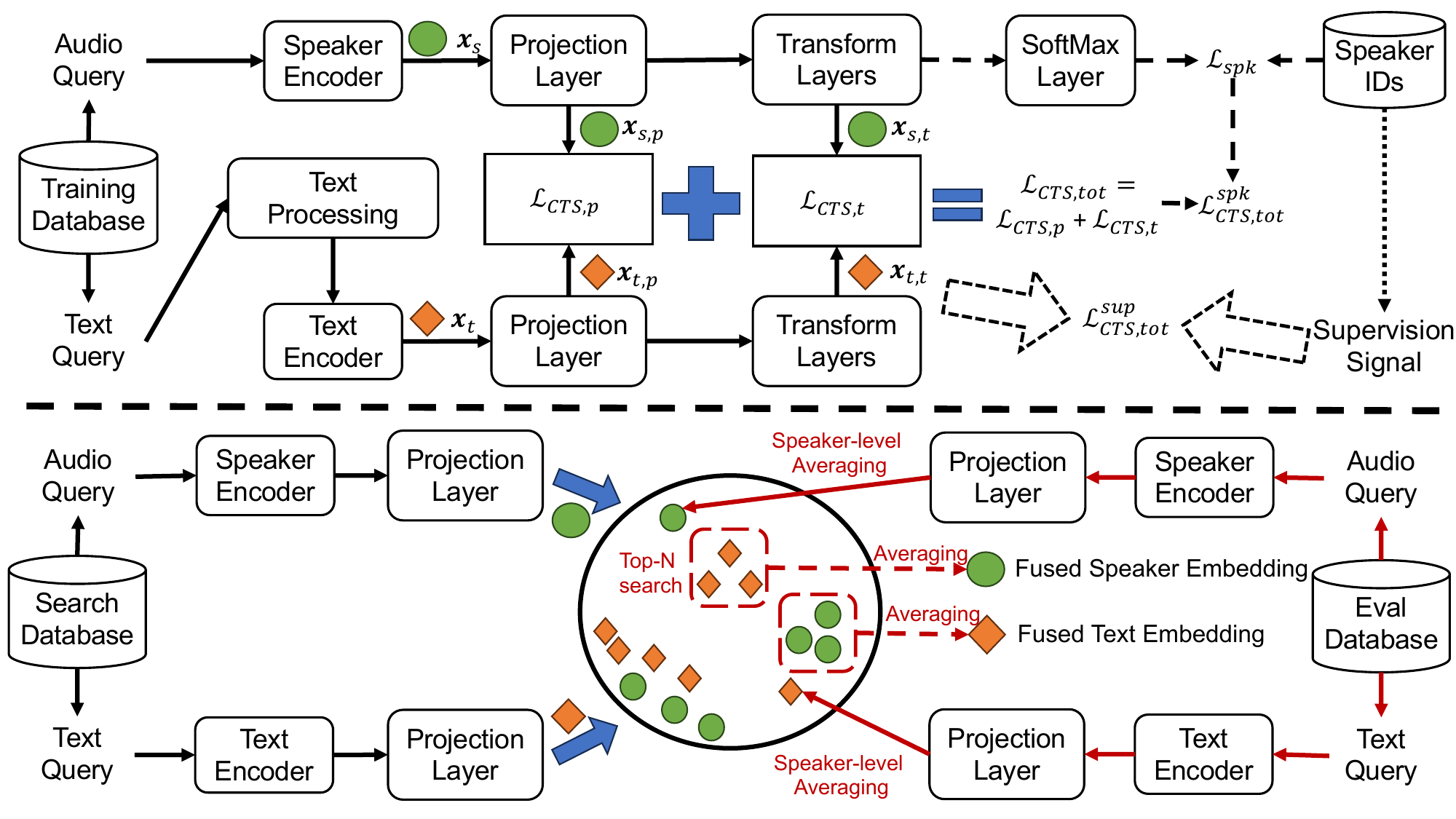}
    \caption{The cross-modal linking framework based on \emph{contrastive learning} (CTS) proposed in this study. The green circles and orange diamonds are speaker and text embeddings, respectively. Best viewed in color.}
    \label{fig:framework}
\vspace{-0.2cm}
\end{figure}

\section{Methodology}
The objective of linking speakers with their associated text is two-fold. First, when presented with an input query regarding speaker characteristics, the goal is to retrieve the most pertinent speaker representation from the provided descriptions. Simultaneously, there is a need to extract the most suitable description of the speaker from the given speaker representation. The speaker representation $\mathbf{x}_{s}$ can be extracted from given audio sample(s), which are fed into a pre-trained speaker encoder. Similarly, the text representation $\mathbf{x}_{t}$ can be extracted from a pre-trained text encoder.

Compared to well-established audio-text retrieval tasks, two notable factors compound the challenge of constructing the search space in this context. Firstly, speaker representations extracted from well-trained speaker encoders \cite{ecapa_tdnn2020} exhibit both speaker-extrinsic and intrinsic variations. Secondly, textual descriptions of speakers, unlike their counterparts in other tasks, lack a comparable level of specificity and hold substantial uncertainties. This preliminary study strategically focus on the foundational establishment of a framework, recognizing it as an initial step targeting these complexities.

\subsection{Linking via Contrastive Learning}
\label{subsec:main_framework}
Our linking framework, depicted in Fig. \ref{fig:framework}, comprises three key stages during the training phase, as shown in the upper section. To begin, input audio and text samples undergo encoding to generate speaker and text embeddings. Text diversity is then enhanced through augmentation and segmentation techniques, as detailed in Section \ref{subsec:text_processing}. Subsequently, these embeddings are passed through a fully-connected (FC) projection layer to ensure dimensionality alignment, followed by further enhancement via multiple FC transformation layers. All representations from the FC layers are length-normalized before proceeding to further steps.

At the training stage, the main loss function included is cross-modal contrastive loss \cite{music_clap2023}. Denoting $i$ as the index of $N$ samples in a minibatch, the backbone of the loss function is expressed as:
\begin{align}
     \mathcal{L}_{s \leftrightarrow t}(\mathbf{x}_{s}, \mathbf{x}_{t}) = -\log( \frac{\exp(\mathbf{x}_{s}^{i} \cdot \mathbf{x}_{t}^{i} / \tau)}{\sum_{j=1}^{N}\exp(\mathbf{x}_{s}^{i} \cdot \mathbf{x}_{t}^{j} / \tau)} )
\end{align}
where $\tau$ is a temperature parameter that can be either fixed or learned. The contrastive loss for the two sets of embeddings can then be calculated as:
\begin{align}
    \mathcal{L}_{CTS}(\mathbf{x}_{s}, \mathbf{x}_{t}) = (\mathcal{L}_{s \leftrightarrow t}(\mathbf{x}_{s}, \mathbf{x}_{t}) + \mathcal{L}_{s \leftrightarrow t}(\mathbf{x}_{t}, \mathbf{x}_{s})) / 2
\label{eq:cts_base_calc}
\end{align}
Based on this formulation, as shown in the figure, we compute this loss via two sets of paired batch embeddings extracted from the projection and transformer layers, respectively: $\mathcal{L}_{CTS, p} = \mathcal{L}_{CTS}(\mathbf{x}_{s,p}, \mathbf{x}_{t,p})$ and $\mathcal{L}_{CTS, t} = \mathcal{L}_{CTS}(\mathbf{x}_{s,t}, \mathbf{x}_{t,t})$. The two losses are summed to form the learning objective:
\begin{align}
    \mathcal{L}_{CTS, tot} = \mathcal{L}_{CTS, p} + \mathcal{L}_{CTS, t}.
\label{eq:cts_calc}
\end{align}

Let us move on to the the inference stage shown at the bottom side of the figure. Firstly, similar to the training database, the search database generates audio and text queries. The embeddings generated from the projection layer ($\mathbf{x}_{s,p}$ and $\mathbf{x}_{s,t}$) are used to form the search space. 

The cross-modal retrieval is realized via the steps shown at the bottom right of the figure, whose transitions are highlighted in red. The projected embedding from the query from one modality is cast into the search space first. The fusion is done by searching for top-$N$ nearest embeddings in the search space from the other modality, followed by averaging over them via distance criteria. The final embedding is averaged from the selected embeddings from the other modality. $N$ is a manually-set parameter. Both the originally projected embeddings and fused embeddings are used for evaluation in Section \ref{sec:results}.

% Depending on the data availability and the practical application scenario, the inference can be discussed in two cases: \emph{Close-set}, where the search database used to build the search space is same as the evaluation query database; \emph{Open-set}, where the search database is different from the evaluation database. Note that for the latter case, the search database may have overlap with 

\subsection{Integrating speaker labels in training}
Since the linking is conducted between speaker and textual representations, it is natural to attempt to incorporate speaker information to improve the performance. As a primer, we employ two ways to include speaker labels into the training.

\textbf{Loss regularization}. First, we employ discriminative speaker classification loss as a regularizer for the main contrastive loss. The classification loss is implemented via \emph{additive angular margin softmax} (AAM-softmax) \cite{aam_softmax_face}, which is widely acquired in state-of-the-art speaker encoder training. The loss is denoted as $\mathcal{L}_{spk}$ in Fig. \ref{fig:framework}. It is used as the regularizer and the new loss is formulated as below:
\begin{align}
    \mathcal{L}_{CTS,tot}^{spk} = \mathcal{L}_{CTS,tot} + \lambda\mathcal{L}_{spk}
\end{align}
where $\lambda$ is the regularization parameter.

\textbf{Supervised contrastive loss}. We regard the speaker labels as a supervision signal for contrastive learning by employing supervised contrastive loss, which was proposed in \cite{supcon2020} and expressed as below:
\begin{align}
   \mathcal{L}_{s \leftrightarrow t}^{sup}(\mathbf{x}_{s}, \mathbf{x}_{t}) = \sum_{i} \frac{-1}{|P(i)|}\sum_{p \in P(i)}\log \frac{\exp(\mathbf{x}_{s}^{i} \cdot \mathbf{x}_{t}^{p} / \tau)}{\sum_{j=1}^{N} \exp(\mathbf{x}_{s}^{i} \cdot \mathbf{x}_{t}^{j} / \tau)} 
\end{align}
where $P(i)$ is the set of all positive pairs in the minibatch and $|P(i)|$ is its cardinality. Such positivity is identified by whether both samples correspond to the same speaker label. The main loss function $\mathcal{L}_{CTS,tot}$, along with its sub-components, are all re-computed accordingly, as denoted in the figure:
\begin{align}
    \mathcal{L}_{CTS,tot}^{sup} = \mathcal{L}_{CTS, p}^{sup} + \mathcal{L}_{CTS, t}^{sup}
\end{align}
where $\mathcal{L}_{CTS, p}^{sup} = (\mathcal{L}_{s \leftrightarrow t}^{sup}(\mathbf{x}_{s,p}, \mathbf{x}_{t,p}) + \mathcal{L}_{s \leftrightarrow t}^{sup}(\mathbf{x}_{t,p}, \mathbf{x}_{s,p})) / 2$ and $\mathcal{L}_{CTS, t}^{sup} = (\mathcal{L}_{s \leftrightarrow t}^{sup}(\mathbf{x}_{s,t}, \mathbf{x}_{t,t}) + \mathcal{L}_{s \leftrightarrow t}^{sup}(\mathbf{x}_{t,t}, \mathbf{x}_{s,t})) / 2$, according to Eq. (\ref{eq:cts_calc}). This method seeks to enhance feature quality in contrastive learning by leveraging supervision from speaker labels, which inherently adds distinctiveness to samples not positively paired (those from different speakers). However, a potential concern is that it could increase the sensitivity of the contrastive learning process to the quality and abundance of the source of the paired data. 
% This is observed and discussed in section \ref{sec:results}.

\begin{figure}
    \centering
    \includegraphics[width=\linewidth]{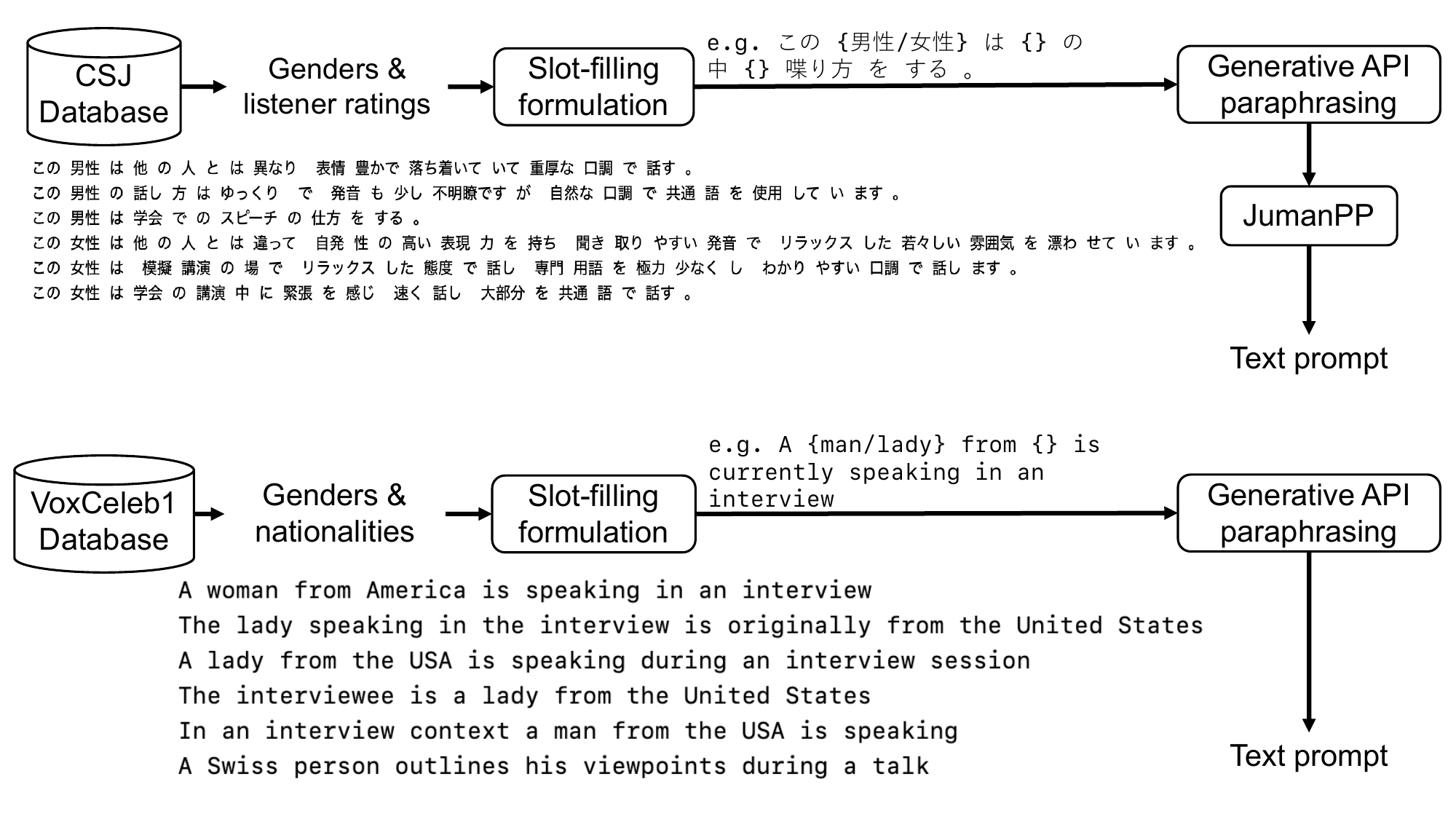}
    \caption{The text processing pipeline for Japanese (up) and English (bottom) scenarios, with sample input texts for each case. Note that for Japanese language, we use \emph{JumanPP} to re-segment the sentence, according to the requirement of the employed text encoder.}
    \label{fig:text_processing}
\vspace{-0.2cm}
\end{figure}

\section{Experimental Setup}
We conduct distinct experiments for both English and Japanese. While the fundamental learning framework remains uniform across both languages, this approach allows us to extract and gain insights that are tailored to the unique characteristics and nuances of each language. This duality in our experimental setup aims at providing a comprehensive understanding of the framework in diverse linguistic contexts.

\subsection{Dataset}
For the English case, we acquire VoxCeleb1 \cite{voxceleb1} for building the protocol. This is a dataset which is sourced from YouTube celebrity interview videos and is widely employed for speaker verification. We follow the protocol described in \cite{voxceleb1} to separate the training and evaluation sets, which contain 1,211 and 40 speakers, respectively.
On the other hand, for the Japanese case, we employ the Corpus of Spontaneous Japanese (CSJ) \cite{csj_corpus}. We also follow the standard protocol outlined in earlier studies \cite{csj_official_asr2015} to select and partition the dataset for our study. The partitioned training data contains 1,388 speakers and 2,672 sessions, covering both simulated and real-time recorded lectures. The evaluation dataset contains 30 speakers. Note that for both cases, following early related works \cite{audio_text2022, msft_clap2023}, the search and evaluation database are both initiated from the evaluation dataset of the corresponding cases.

There are two important differences between these two scenarios. First, in terms of audio quality, VoxCeleb1 is more noisy and uncontrolled than CSJ. This is expected due to their sources. Another difference is the fruitfulness of text --- as detailed in Section \ref{subsec:text_processing} and illustrated in Fig. \ref{fig:text_processing}, due to the amount of metadata available, following the same pipeline, Japanese input texts are more fruitful and varied compared to English.

\subsection{Models}
\textbf{Speaker encoders}. We use ECAPA-TDNN \cite{ecapa_tdnn2020} as the input speaker encoder for our baseline model. We use the implementation from SpeechBrain\footnote{\url{https://huggingface.co/speechbrain/spkrec-ecapa-voxceleb}}. The audio sampling frequency is 16 kHz. For comparison, we also test general-purpose audio encoders: PANN\footnote{\url{https://github.com/qiuqiangkong/panns_inference}} \cite{pann} Wav2Vec2.0\footnote{\url{https://huggingface.co/facebook/wav2vec2-large-960h}}\footnote{The features are extracted from the feature encoder of the network, as defined in \cite{wav2vec2}.} \cite{wav2vec2}.
Note that the sampling frequency of the used pre-trained PANN is 32 kHz, so up-sampling is done.

\textbf{Text encoders}. Different pre-trained language models for the two languages covered in this study are acquired. For English scenario, we use BART \cite{bert} sourced from Hugging Face\footnote{\url{https://huggingface.co/bert-base-uncased}}. For the Japanese scenario, we acquire an implementation of the same language model, adopted and trained on Japanese Wikipedia\footnote{\url{https://huggingface.co/ku-nlp/bart-large-japanese}}. The dimension of text embeddings generated are 768 and 1,024, respectively.

\textbf{Network and training}. The input dimension of projected speaker embeddings from speaker and text encoders are projected to 768 dimensions by the projection layer. The width of the transform layers is also 768. Various loss functions are used for training the network, as presented in Section \ref{subsec:main_framework}. When including the regularization term, $\lambda=0.1$.

\textbf{Cross-modal fusion}. During the inference phase, the search space employs cosine similarity \cite{cosine_similarity} as the fundamental metric. When searching for the top-\textit{N} nearest embeddings, $N=10$. A similar fusion procedure is executed for unlinked embeddings, wherein the textual embeddings undergo dimensionality alignment with the speaker embeddings via the application of linear discriminant analysis (LDA) \cite{lda}.

\subsection{Text processing}
\label{subsec:text_processing}
% Different from other cross-modal learning tasks, this specific task does not usually come with well-annotated paired data. Instead, speaker-related information is commonly manifested in the guise of metadata, a resource that can be either copious or scarce. Users also often anticipate the ability to construct structural prompts conveniently, obviating the need for extensive human annotation efforts. Moreover, specialized text encoders may adhere to established conventions tailored to specific languages. Therefore, the prompt building method applied is contingent on the availability of the metadata. 

Fig. \ref{fig:text_processing} demonstrates the text processing pipeline for the two languages. In order to avoid extensive human annotation efforts, for both languages, there are certain steps needed: 1) Obtain the speaker-related metadata from the database; 2) Perform a slot-filling formulation (via the brackets in the figure) to construct structured text prompts for speakers; 3) Consult a large language model\footnote{\url{https://platform.openai.com/docs/models/gpt-3-5}} via user interfaces to paraphrase the structured prompts, creating variations across speakers\footnote{We used \emph{equal error rate} (EER) on text prompts for the speakers using the corresponding text encoders, to measure the variations. Reported text EERs are 1.03\% and 0\% for English and Japanese cases, respectively.}. Note that compared to the English scenario, for the Japanese case, there is an extra re-segmentation step to create the final prompt, using the library \emph{JumanPP} \cite{jumanpp2015}\footnote{\url{https://github.com/ku-nlp/jumanpp}}. It relies on recommendations from the text encoder pipeline utilized in the Japanese scenario.

\subsection{Evaluation metrics}
\label{subsec:eval_metrics}
We employ two complementary and independent metrics for results, \emph{mean average precision} (mAP) and mean rank (MeanR), for assessing the performance in speaker-to-text ($s \rightarrow t$) and text-to-speaker retrieval ($t \rightarrow s$). These metrics are extensively utilized in the context of cross-modal retrieval tasks, as substantiated by previous research \cite{video_audio_retrieval2019, audio_text2022}. A higher mAP value signifies more precise retrieval across multiple queries, while a lower MeanR value indicates that the retrieval target has a better chance of being ranked highly. Furthermore, to provide additional insights, we offer a visualization of the embeddings through t-SNE \cite{tsne}, which aids in examining the quality of clustering and its correlation with retrieval.

\begin{table*}[ht]
\scriptsize
    \centering
    \caption{English cross-modal retrieval results. The text encoder used in \cite{msft_clap2023} is RoBERTa \cite{liu2019roberta}.}
    \begin{tabular}{|c|c|c|cc|cc|cc|cc|}
    \hline
        & & & \multicolumn{2}{|c|}{$s \rightarrow t$} & \multicolumn{2}{|c|}{$t \rightarrow s$} & \multicolumn{2}{|c|}{$s \rightarrow t$(fusion)} & \multicolumn{2}{|c|}{$t \rightarrow s$(fusion)} \\ \hline
        System & Speaker/Audio Enc. & Loss & mAP10$\uparrow$ & MeanR$\downarrow$ & mAP10$\uparrow$ & MeanR$\downarrow$ & mAP10$\uparrow$ & MeanR$\downarrow$ & mAP10$\uparrow$ & MeanR$\downarrow$ \\ \hline
        \cite{msft_clap2023} & \cite{htsat} & - & \phantom{0}7.32 & 21.02 & \phantom{0}7.32 & 21.65 & \phantom{0}6.93 & 20.88 & \phantom{0}7.29 & 20.52 \\ \hline
        \cite{msft_clap2023} (fine-tuned) & \cite{htsat} & - & 15.78 & 12.48 & 18.14 & \textbf{10.92} & 15.12 & 12.30 & 13.20 & 14.85  \\ \hline
        \texttt{B00} & ECAPA-TDNN & (unlinked) & \phantom{0}7.58 & 18.75 & \phantom{0}8.86 & 19.95 & \phantom{0}6.15 & 19.98 & 10.36 & 19.92 \\ \hline \hline
        \texttt{B01} & PANN & $\mathcal{L}_{CTS,tot}$ & 16.39 & \textbf{10.80} & 16.62 & 12.08 & 17.30 & \phantom{0}9.88 & 18.59 & \phantom{0}9.80 \\ \hline
        \texttt{B02} & Wav2Vec2.0 & $\mathcal{L}_{CTS,tot}$ & 18.47 & 12.10 & 14.62 & 13.05 & 18.68 & 10.02 & \textbf{24.39} & \phantom{0}\textbf{8.88} \\ \hline
        \texttt{B03} & ECAPA-TDNN & $\mathcal{L}_{CTS,tot}$ & 18.61 & 12.00 & \textbf{19.97} & 13.52 & 23.00 & \phantom{0}9.22 & 21.86 & \phantom{0}9.43 \\ \hline\hline
        \texttt{C01} & ECAPA-TDNN & $\mathcal{L}_{CTS,tot}^{spk}$ & \textbf{22.79} & 11.68 & 19.48 & 12.25 & 19.14 & \phantom{0}9.85 & 15.95 & \phantom{0}11.0 \\ \hline
        \texttt{C02} & ECAPA-TDNN & $\mathcal{L}_{CTS,tot}^{supcon}$ & 19.29 & 11.60 & 17.74 & 12.45 & \textbf{23.58} & \phantom{0}\textbf{8.45} & 20.89 & \phantom{0}9.88 \\ \hline
    \end{tabular}
    \label{tab:main_results}
\end{table*}

\begin{table*}[ht]
\scriptsize
    \centering
    \caption{Japanese cross-modal retrieval results.}
    \begin{tabular}{|c|c|c|cc|cc|cc|cc|}
    \hline
        & & & \multicolumn{2}{|c|}{$s \rightarrow t$} & \multicolumn{2}{|c|}{$t \rightarrow s$} & \multicolumn{2}{|c|}{$s \rightarrow t$(fusion)} & \multicolumn{2}{|c|}{$t \rightarrow s$(fusion)} \\ \hline
        System & Speaker/Audio Enc. & Loss & mAP10$\uparrow$ & MeanR$\downarrow$ & mAP10$\uparrow$ & MeanR$\downarrow$ & mAP10$\uparrow$ & MeanR$\downarrow$ & mAP10$\uparrow$ & MeanR$\downarrow$ \\ \hline
        \texttt{B00} & ECAPA-TDNN & (unlinked) & 11.53 & 16.20 & \phantom{0}6.72 & 16.07 & \phantom{0}7.19 & 15.10 & \phantom{0}8.39 & 16.40 \\ \hline \hline
        \texttt{B01} & PANN & $\mathcal{L}_{CTS,tot}$ & 38.54 & \phantom{0}6.90 & 35.09 & \phantom{0}6.50 & 34.70 & \phantom{0}5.27 & \textbf{43.09} & \phantom{0}5.63   \\ \hline
        \texttt{B02} & Wav2Vec2.0 & $\mathcal{L}_{CTS,tot}$ & 36.94 & \phantom{0}\textbf{6.23} & 33.57 & \phantom{0}\textbf{6.03} & \textbf{42.02} & \phantom{0}\textbf{4.67} & 41.78 & \phantom{0}\textbf{4.43}  \\ \hline
        \texttt{B03} & ECAPA-TDNN & $\mathcal{L}_{CTS,tot}$ & \textbf{40.93} & \phantom{0}\textbf{6.23} & \textbf{43.17} & \phantom{0}6.73 & 34.90 & \phantom{0}5.63 & 36.81 & \phantom{0}5.63  \\ \hline\hline
        \texttt{C01} & ECAPA-TDNN & $\mathcal{L}_{CTS,tot}^{spk}$ & 26.43 & \phantom{0}9.07 & 27.70 & \phantom{0}8.67 & 33.67 & \phantom{0}5.90 & 33.58 & \phantom{0}6.90 \\ \hline
        \texttt{C02} & ECAPA-TDNN & $\mathcal{L}_{CTS,tot}^{supcon}$ & 27.92 & \phantom{0}9.67 & 24.27 & 10.53 & 26.12 & \phantom{0}9.17 & 24.34 & \phantom{0}8.00 \\ \hline
    \end{tabular}
    \label{tab:main_results}
    \vspace{-0.3cm}
\end{table*}

\section{Results}
\label{sec:results}
The statistical results are detailed in Table \ref{tab:main_results}. We conduct comparative analysis between several systems for the English scenario: one where embeddings are directly extracted from the encoders without any linking (referred to as ``unlinked", \texttt{B00}), a previous study \cite{msft_clap2023} that integrated audio transformers \cite{htsat} for event classification, its fine-tuned variant\footnote{The fine-tuning is done by additional training using VoxCeleb1 training data with binary cross-entropy loss, based on pilot experiments.}, along with our proposed ones. The scope of this comparison is constrained to the English scenario, as the comparison systems were exclusively trained on English data. The results in both metrics indicate that our proposed baseline model (\texttt{B03}) demonstrated improvements in mAP10, showcasing its ability to accurately locate target embeddings from both text-to-speaker and speaker-to-text perspectives, outperforming the fine-tuned pre-trained model in MeanR for text-to-speaker and speaker-to-text retrieval using fused embeddings.

Subsequently, we examine the impact of employing alternative audio encoders via systems, trained on massive data for different tasks, compared to the speaker encoder. For English, ECAPA-TDNN demonstrates marginally superior performance in mAP10 in contrast to other audio encoders, with the notable exception of text-to-speaker retrieval using fused embeddings, where it is slightly surpassed by Wav2Vec2.0 in both instances. This discrepancy may be attributed to the augmented information assimilated into the text embeddings via the speaker learning branch, which does not necessarily pertain exclusively to speakers. A parallel observation is noted in the case of Japanese, with a more pronounced performance differential, potentially attributable to the richer contextual information present in the text branch, echoing more effectively with the speaker branch.

Moving on to systems with the prefix \texttt{C}, our investigation reveals that the inclusion of speaker labels during training yields distinct results for English and Japanese. In the Japanese context, incorporating speaker labels leads to a decrease in performance on both metrics, suggesting that speaker labels may have a detrimental impact when integrated in the proposed manner for effective prompts. For English, while including speaker labels as a regularizer does not consistently improve performance, consistent enhancements are observed when using speaker labels to supervise training at the minibatch level. This suggests that in scenarios with more challenging data (noisier audio, less structured text input), incorporating speaker labels as additional information can be beneficial. This observation is particularly interesting as it reflects real-time user expectations, which may vary depending on the data characteristics.

\begin{figure}[t]
    \centering
    \subfloat[Speaker embedding $\mathbf{x}_{s,p}$\label{fig7:audio}]{
    \includegraphics[width=0.5\linewidth]{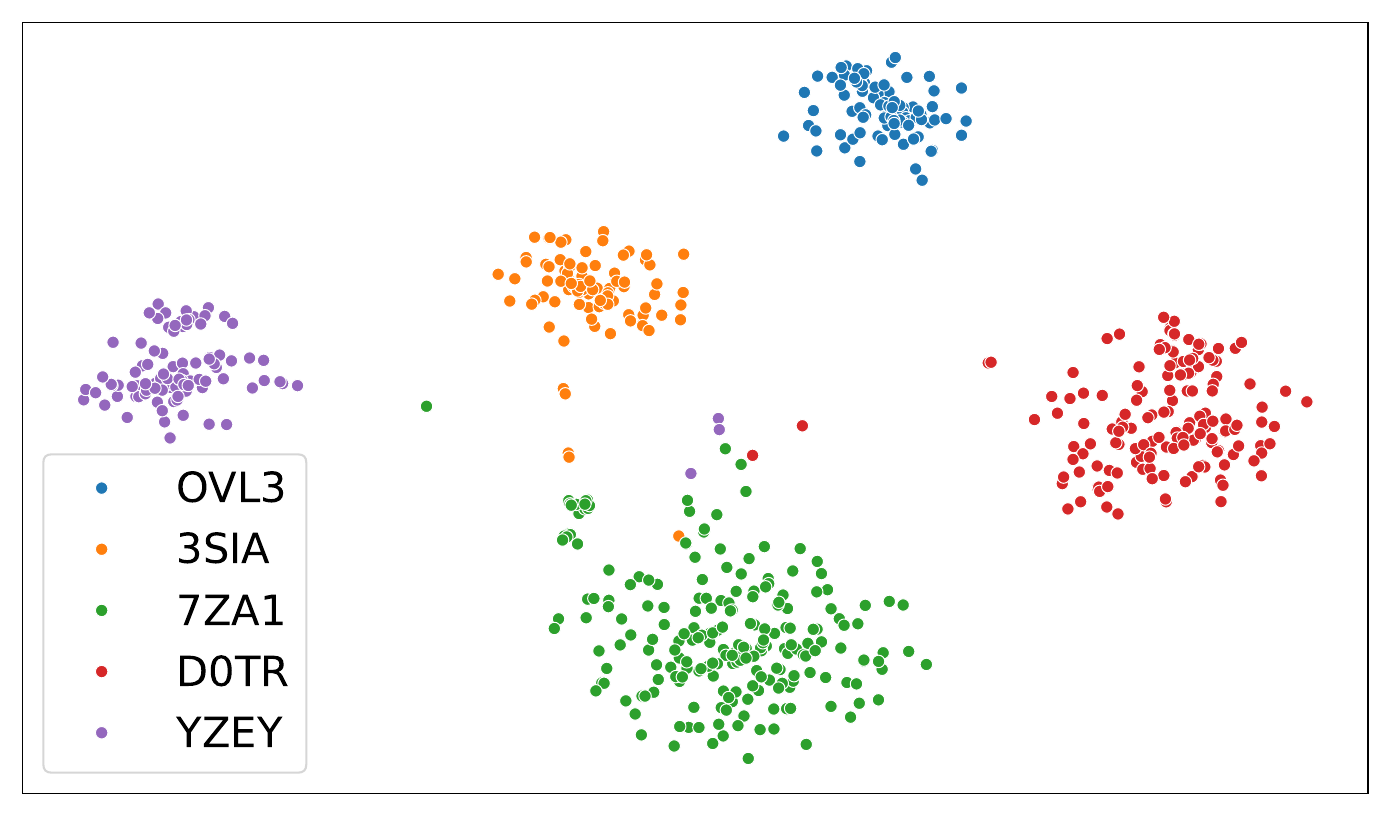} 
    }
    \subfloat[Text embedding $\mathbf{x}_{t,p}$\label{fig7:text}]{
    \includegraphics[width=0.5\linewidth]{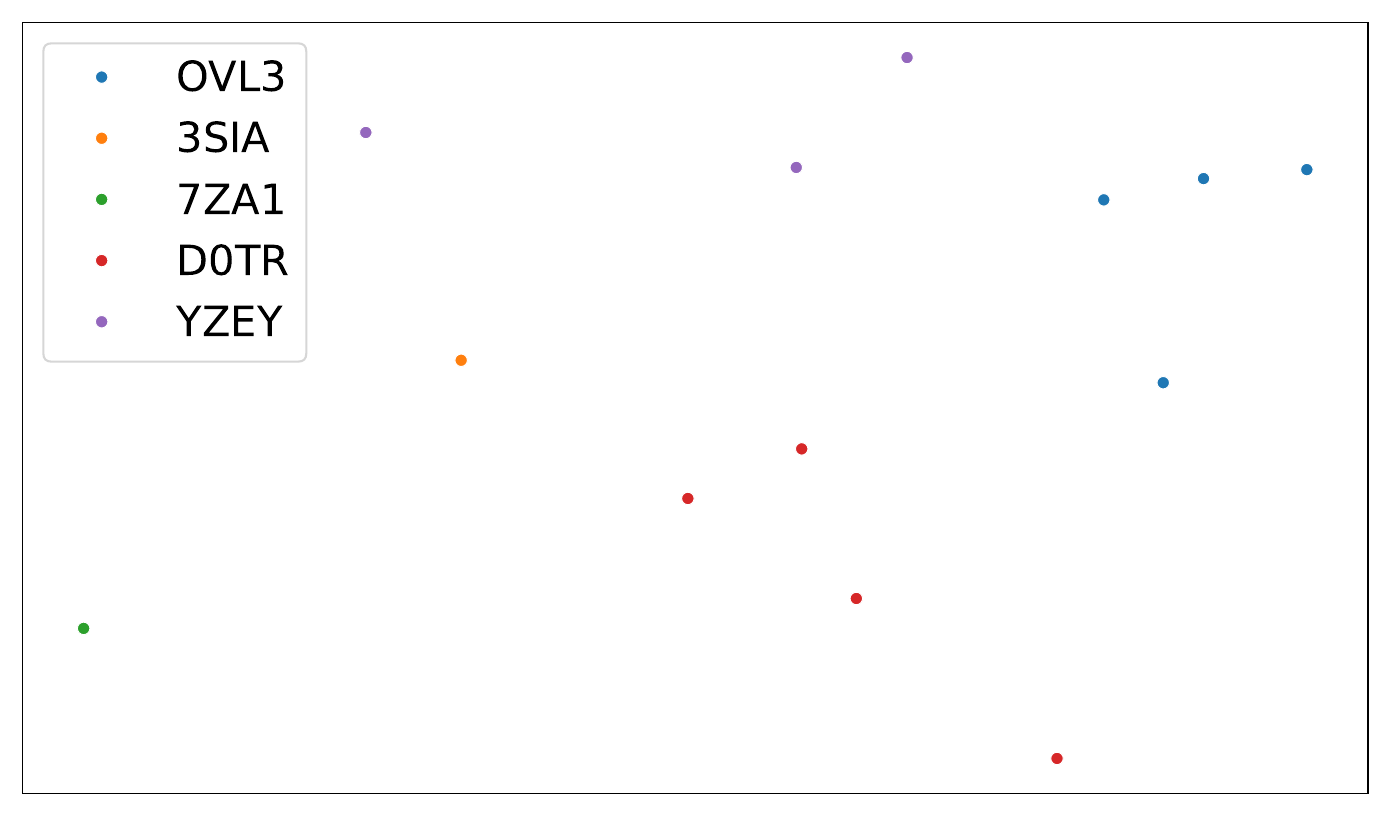} 
    } \par
    \subfloat[Speaker embedding $\mathbf{x}_{s,p}$ (fused)\label{fig7:infer_audio}]{
    \includegraphics[width=0.5\linewidth]{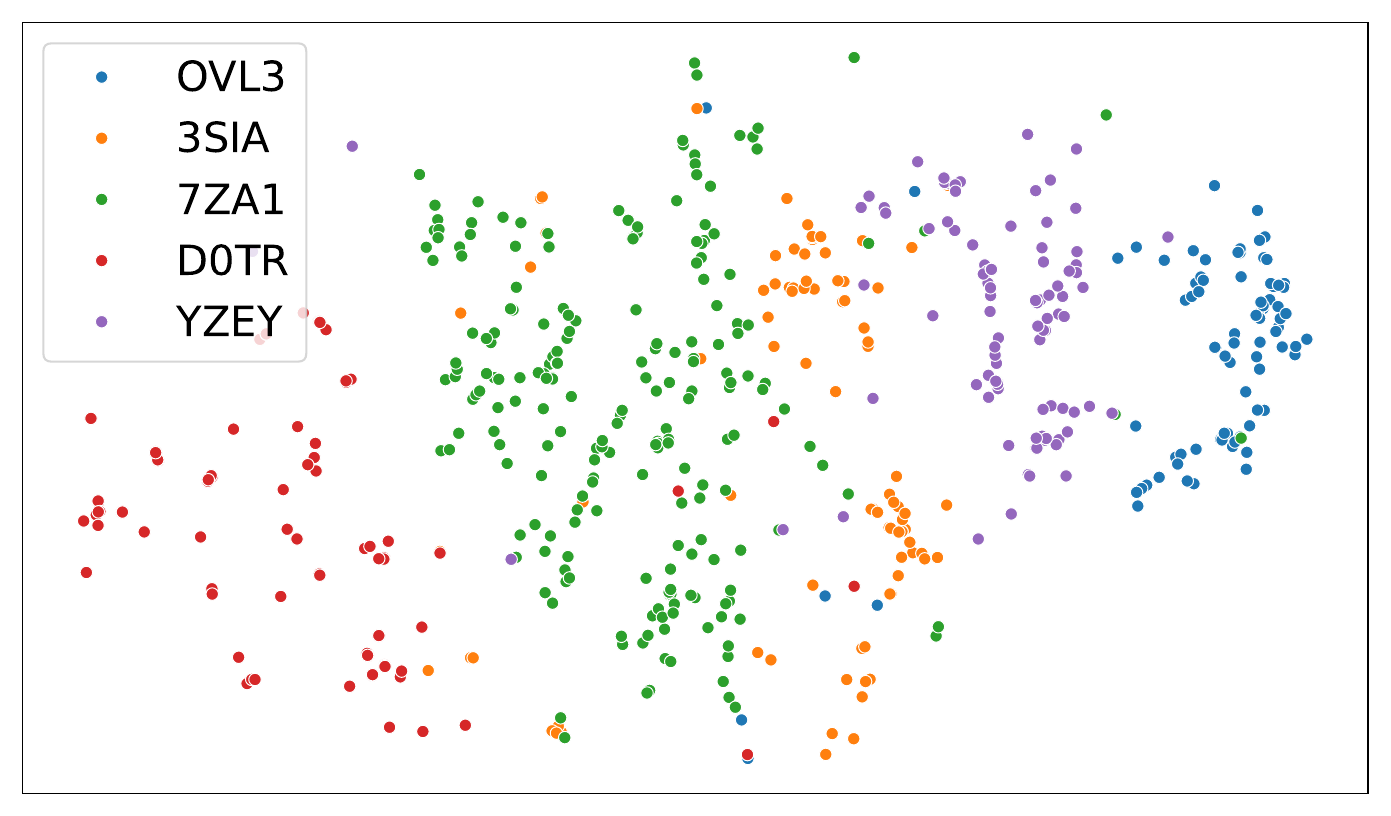} 
    }
    \subfloat[Text embedding $\mathbf{x}_{t,p}$ (fused)\label{fig7:infer_text}]{
    \includegraphics[width=0.5\linewidth]{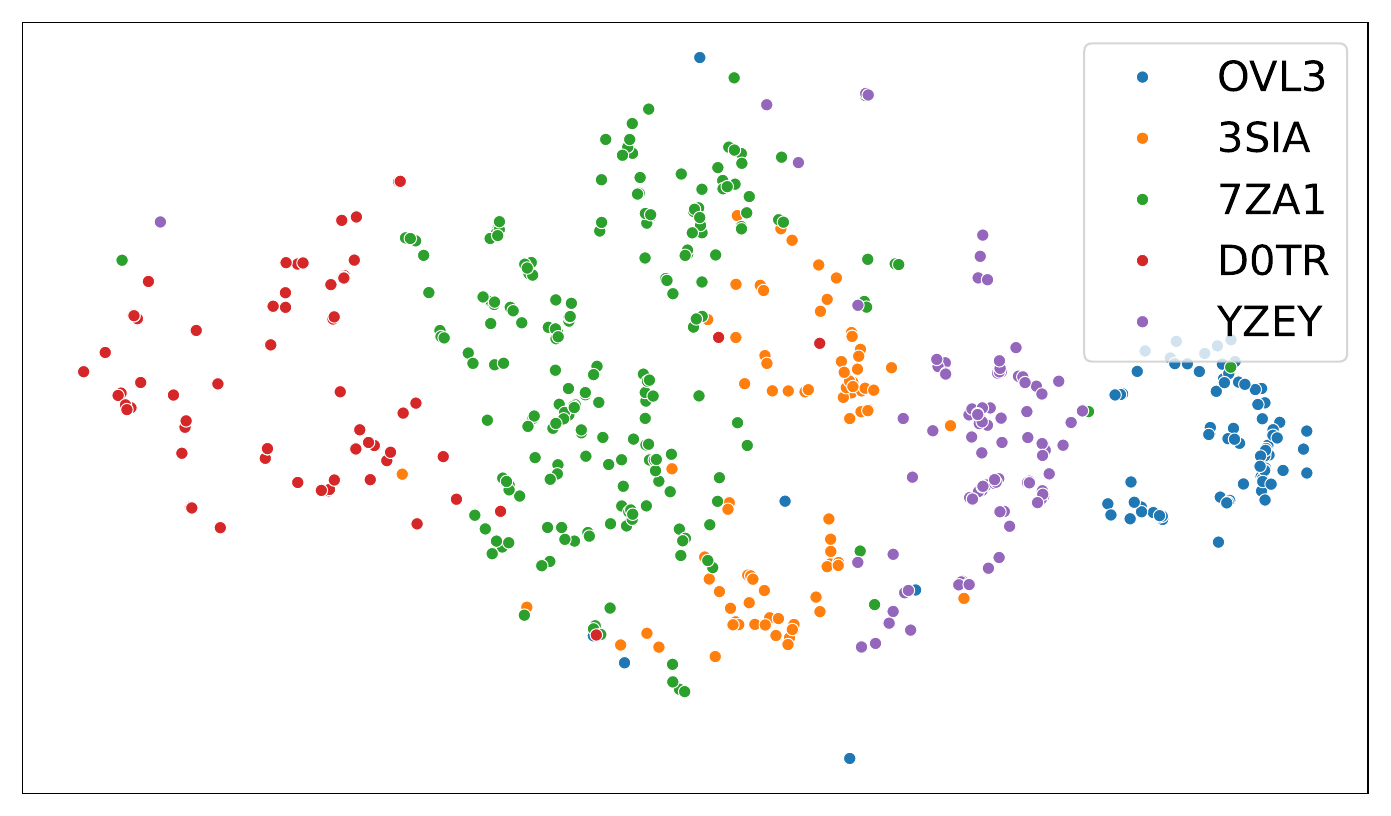} 
    } 
    \caption{t-SNE visualizations of extracted embeddings and their fused counterparts from another modal at inference stage. The speakers are anonymized in the figure. Best viewed in color.}
    \label{fig:visuals}
\vspace{-0.2cm}
\end{figure}

Lastly, as depicted in Fig. \ref{fig:visuals}, we present speaker clustering performance using different types of embeddings computed from \texttt{B03}. For clarity, five speakers are randomly selected from the CSJ evaluation set and their identities are anonymized. The visual representation highlights that while cluster patterns are roughly maintained during fusion with another modality, there is an increased overlap between speakers compared to their original representations. Correlating this observation with the results in Table \ref{tab:main_results}, it is evident that such a level of overlap among speaker representations contributes to improvements in in-list retrieval performance, particularly in the English scenario. The impact is marginally less consistent in the Japanese scenario, which is possibly attributable to cleaner acoustic conditions and richer textual information.

\section{Conclusion}
In this study, we have initiated an exploration of linking speaker-related and textual data using contrastive learning to facilitate fine-grained cross-modal retrieval. Our proposed framework involves well-defined training and inference phases, employing straightforward neural architectures and cosine similarity, with the optional involvement of speaker labels. Our findings highlight enhanced retrieval performance from the proposed methods compared to unlinked cases, with attempts on alternatives apart from the speaker encoder for encoding speaker information. Visual analysis underscores the utility of increased overlap in speaker representations achieved through cross-modal fusion. Future research may focus on developing efficient methods for incorporating speaker information into speaker-text retrieval.

\bibliographystyle{IEEEtran}
\bibliography{refs}

\begin{thebibliography}{10}
\providecommand{\url}[1]{#1}
\csname url@samestyle\endcsname
\providecommand{\newblock}{\relax}
\providecommand{\bibinfo}[2]{#2}
\providecommand{\BIBentrySTDinterwordspacing}{\spaceskip=0pt\relax}
\providecommand{\BIBentryALTinterwordstretchfactor}{4}
\providecommand{\BIBentryALTinterwordspacing}{\spaceskip=\fontdimen2\font plus
\BIBentryALTinterwordstretchfactor\fontdimen3\font minus \fontdimen4\font\relax}
\providecommand{\BIBforeignlanguage}[2]{{%
\expandafter\ifx\csname l@#1\endcsname\relax
\typeout{** WARNING: IEEEtran.bst: No hyphenation pattern has been}%
\typeout{** loaded for the language `#1'. Using the pattern for}%
\typeout{** the default language instead.}%
\else
\language=\csname l@#1\endcsname
\fi
#2}}
\providecommand{\BIBdecl}{\relax}
\BIBdecl

\bibitem{image_text_alignment2015}
A.~Karpathy and L.~Fei-Fei, ``Deep visual-semantic alignments for generating image descriptions,'' \emph{IEEE Transactions on Pattern Analysis and Machine Intelligence}, vol.~39, no.~4, pp. 664--676, 2017.

\bibitem{clip}
A.~Radford, J.~W. Kim, C.~Hallacy, A.~Ramesh, G.~Goh, S.~Agarwal, G.~Sastry, A.~Askell, P.~Mishkin, J.~Clark, G.~Krueger, and I.~Sutskever, ``Learning transferable visual models from natural language supervision,'' in \emph{Proceedings of the 38th International Conference on Machine Learning}, ser. Proceedings of Machine Learning Research, M.~Meila and T.~Zhang, Eds., vol. 139.\hskip 1em plus 0.5em minus 0.4em\relax PMLR, 18--24 Jul 2021, pp. 8748--8763.

\bibitem{video_text_retrieval2015}
R.~Xu, C.~Xiong, W.~Chen, and J.~J. Corso, ``Jointly modeling deep video and compositional text to bridge vision and language in a unified framework,'' in \emph{Proceedings of the Twenty-Ninth AAAI Conference on Artificial Intelligence}, ser. AAAI'15.\hskip 1em plus 0.5em minus 0.4em\relax AAAI Press, 2015, p. 2346–2352.

\bibitem{video_text_retrieval2019}
Y.~Liu, S.~Albanie, A.~Nagrani, and A.~Zisserman, ``Use what you have: Video retrieval using representations from collaborative experts,'' in \emph{British Machine Vision Conference (BMVC)}.

\bibitem{video_audio_retrieval2019}
D.~Sur\'{\i}s, A.~Duarte, A.~Salvador, J.~Torres, and X.~Gir\'{o}-i Nieto, ``Cross-modal embeddings for video and audio retrieval,'' in \emph{Computer Vision – ECCV 2018}, 2019, p. 711–716.

\bibitem{audio_text_retrieval2008}
\BIBentryALTinterwordspacing
G.~Chechik, E.~Ie, M.~Rehn, S.~Bengio, and D.~Lyon, ``Large-scale content-based audio retrieval from text queries,'' in \emph{Proceedings of the 1st ACM International Conference on Multimedia Information Retrieval}, ser. MIR '08.\hskip 1em plus 0.5em minus 0.4em\relax New York, NY, USA: Association for Computing Machinery, 2008, p. 105–112. [Online]. Available: \url{https://doi.org/10.1145/1460096.1460115}
\BIBentrySTDinterwordspacing

\bibitem{siamese2016}
M.~Cen and C.~Jung, ``Fully convolutional siamese fusion networks for object tracking,'' in \emph{2018 25th IEEE International Conference on Image Processing (ICIP)}, 2018, pp. 3718--3722.

\bibitem{msft_clap2023}
B.~Elizalde, S.~Deshmukh, M.~A. Ismail, and H.~Wang, ``Clap learning audio concepts from natural language supervision,'' in \emph{Proc. ICASSP}, 2023, pp. 1--5.

\bibitem{laion_clap2023}
Y.~Wu, K.~Chen, T.~Zhang, Y.~Hui, T.~Berg-Kirkpatrick, and S.~Dubnov, ``Large-scale contrastive language-audio pretraining with feature fusion and keyword-to-caption augmentation,'' in \emph{Proc. ICASSP}, 2023, pp. 1--5.

\bibitem{audioset}
J.~F. Gemmeke, D.~P.~W. Ellis, D.~Freedman, A.~Jansen, W.~Lawrence, R.~C. Moore, M.~Plakal, and M.~Ritter, ``Audio set: An ontology and human-labeled dataset for audio events,'' in \emph{Proc. ICASSP}, 2017, pp. 776--780.

\bibitem{clotho}
K.~Drossos, S.~Lipping, and T.~Virtanen, ``Clotho: an audio captioning dataset,'' in \emph{Proc. ICASSP}, 2020, pp. 736--740.

\bibitem{pann}
Q.~Kong, Y.~Cao, T.~Iqbal, Y.~Wang, W.~Wang, and M.~D. Plumbley, ``Panns: Large-scale pretrained audio neural networks for audio pattern recognition,'' \emph{IEEE/ACM Transactions on Audio, Speech, and Language Processing}, vol.~28, pp. 2880--2894, 2020.

\bibitem{htsat}
K.~Chen, X.~Du, B.~Zhu, Z.~Ma, T.~Berg-Kirkpatrick, and S.~Dubnov, ``{HTS-AT}: A hierarchical token-semantic audio transformer for sound classification and detection,'' in \emph{Proc. ICASSP}, 2022, pp. 646--650.

\bibitem{bert}
\BIBentryALTinterwordspacing
J.~Devlin, M.-W. Chang, K.~Lee, and K.~Toutanova, ``{BERT}: Pre-training of deep bidirectional transformers for language understanding,'' in \emph{Proceedings of the 2019 Conference of the North {A}merican Chapter of the Association for Computational Linguistics: Human Language Technologies, Volume 1 (Long and Short Papers)}.\hskip 1em plus 0.5em minus 0.4em\relax Minneapolis, Minnesota: Association for Computational Linguistics, Jun. 2019, pp. 4171--4186. [Online]. Available: \url{https://aclanthology.org/N19-1423}
\BIBentrySTDinterwordspacing

\bibitem{probe_xvector2019}
D.~Raj, D.~Snyder, D.~Povey, and S.~Khudanpur, ``Probing the information encoded in x-vectors,'' in \emph{2019 IEEE Automatic Speech Recognition and Understanding Workshop (ASRU)}, 2019, pp. 726--733.

\bibitem{ecapa_tdnn2020}
B.~Desplanques, J.~Thienpondt, and K.~Demuynck, ``{ECAPA-TDNN}: Emphasized channel attention, propagation and aggregation in tdnn based speaker verification,'' in \emph{Proc. Interspeech}, 2020, pp. 3830--3834.

\bibitem{music_clap2023}
S.~Doh, M.~Won, K.~Choi, and J.~Nam, ``Toward universal text-to-music retrieval,'' in \emph{Proc. ICASSP}, 2023, pp. 1--5.

\bibitem{aam_softmax_face}
F.~Wang, J.~Cheng, W.~Liu, and H.~Liu, ``Additive margin softmax for face verification,'' \emph{IEEE Signal Processing Letters}, vol.~25, no.~7, pp. 926--930, 2018.

\bibitem{supcon2020}
P.~Khosla, P.~Teterwak, C.~Wang, A.~Sarna, Y.~Tian, P.~Isola, A.~Maschinot, C.~Liu, and D.~Krishnan, ``Supervised contrastive learning,'' ser. NIPS'20.\hskip 1em plus 0.5em minus 0.4em\relax Red Hook, NY, USA: Curran Associates Inc., 2020.

\bibitem{voxceleb1}
A.~Nagrani, J.~Chung, and A.~Zisserman, ``{VoxCeleb}: A large-scale speaker identification dataset,'' in \emph{Proc. Interspeech}, 2017, pp. 2616--2620.

\bibitem{csj_corpus}
K.~Maekawa, ``{Corpus of spontaneous Japanese: its design and evaluation},'' in \emph{Proc. ISCA/IEEE Workshop on Spontaneous Speech Processing and Recognition}, 2003, p. paper MMO2.

\bibitem{csj_official_asr2015}
T.~Moriya, T.~Tanaka, T.~Shinozaki, S.~Watanabe, and K.~Duh, ``Automation of system building for state-of-the-art large vocabulary speech recognition using evolution strategy,'' in \emph{2015 IEEE Workshop on Automatic Speech Recognition and Understanding (ASRU)}, 2015, pp. 610--616.

\bibitem{audio_text2022}
S.~Lou, X.~Xu, M.~Wu, and K.~Yu, ``Audio-text retrieval in context,'' in \emph{Proc. ICASSP}, 2022, pp. 4793--4797.

\bibitem{wav2vec2}
A.~Baevski, H.~Zhou, A.~Mohamed, and M.~Auli, ``Wav2vec 2.0: A framework for self-supervised learning of speech representations,'' in \emph{Proceedings of the 34th International Conference on Neural Information Processing Systems}, ser. NIPS'20.\hskip 1em plus 0.5em minus 0.4em\relax Red Hook, NY, USA: Curran Associates Inc., 2020.

\bibitem{cosine_similarity}
A.~Singhal, ``Modern information retrieval: A brief overview.'' \emph{IEEE Data Eng. Bull.}, vol.~24, no.~4, pp. 35--43, 2001.

\bibitem{lda}
K.~Fukunaga, \emph{Introduction to Statistical Pattern Recognition}, 2nd~ed.\hskip 1em plus 0.5em minus 0.4em\relax Academic Press, 1990.

\bibitem{jumanpp2015}
\BIBentryALTinterwordspacing
H.~Morita, D.~Kawahara, and S.~Kurohashi, ``Morphological analysis for unsegmented languages using recurrent neural network language model,'' in \emph{Proceedings of the 2015 Conference on Empirical Methods in Natural Language Processing}, L.~M{\`a}rquez, C.~Callison-Burch, and J.~Su, Eds.\hskip 1em plus 0.5em minus 0.4em\relax Lisbon, Portugal: Association for Computational Linguistics, Sep. 2015, pp. 2292--2297. [Online]. Available: \url{https://aclanthology.org/D15-1276}
\BIBentrySTDinterwordspacing

\bibitem{tsne}
\BIBentryALTinterwordspacing
L.~van~der Maaten and G.~Hinton, ``Visualizing data using t-sne,'' \emph{Journal of Machine Learning Research}, vol.~9, no.~86, pp. 2579--2605, 2008. [Online]. Available: \url{http://jmlr.org/papers/v9/vandermaaten08a.html}
\BIBentrySTDinterwordspacing

\bibitem{liu2019roberta}
\BIBentryALTinterwordspacing
Y.~Liu, M.~Ott, N.~Goyal, J.~Du, M.~Joshi, D.~Chen, O.~Levy, M.~Lewis, L.~Zettlemoyer, and V.~Stoyanov, ``Roberta: A robustly optimized bert pretraining approach,'' 2019, cite arxiv:1907.11692. [Online]. Available: \url{http://arxiv.org/abs/1907.11692}
\BIBentrySTDinterwordspacing

\end{thebibliography}

\end{document}